\documentclass[draftcls, onecolumn, 12pt]{IEEEtran}
\usepackage{amsfonts,amssymb,amsmath,bm}
\usepackage{graphicx,subfigure}
\usepackage{cite,color}
\usepackage{booktabs}
\usepackage{threeparttable}
\usepackage{mathtools}

\newtheorem{prop}{Proposition}

\ifCLASSINFOpdf
\else
\fi
\allowdisplaybreaks[0]

\begin{document}
%
\title{Distributed Optimal Rate-Reliability-Lifetime Tradeoff in Wireless Sensor Networks}

\author{ Weiqiang~Xu, 
Qingjiang~Shi, 
Xiaoyun Wei,
Yaming Wang

\thanks{W. Xu, Q. Shi, X. Wei, Y. Wang
are with School of Information Science \& Technology, Zhejiang Sci-Tech University, Hangzhou, 310018, P. R. China. (email: wqxu@zstu.edu.cn).

}
}

\maketitle

\begin{abstract}
The transmission rate, delivery reliability and network lifetime are three fundamental but conflicting design objectives in energy-constrained wireless sensor networks.
In this paper, we address the optimal rate-reliability-lifetime tradeoff with link capacity constraint, reliability constraint and energy constraint. By introducing the weight parameters, we combine the objectives at rate, reliability, and lifetime into a single objective to characterize the tradeoff among them. However, the optimization formulation of the rate-reliability-reliability tradeoff is neither separable nor convex. Through a series of transformations, a separable and convex problem is derived, and an efficient distributed Subgradient Dual Decomposition algorithm (SDD) is proposed.
Numerical examples confirm its convergence.
Also, numerical examples investigate the impact of weight parameters on the rate utility, reliability utility and network lifetime, which provide a guidance to properly set the value of weight parameters for a desired performance of WSNs according to the realistic  application's requirements.
\end{abstract}

\begin{IEEEkeywords}
Wireless sensor network, network utility maximization, rate allocation, reliability, network lifetime maximization .
\end{IEEEkeywords}

%
\IEEEpeerreviewmaketitle

\newpage
\section{Introduction}
\IEEEPARstart{W}{ireless} Sensor Networks (WSNs) consist of a
large number of spatially distributed autonomous sensor
nodes with limited
computation and sensing capabilities,
 to monitor physical or environmental conditions, and to cooperatively pass their data to a sink.
They have been extensively applied in many fields, such as battlefield surveillance, environmental monitoring, home automation, critical infrastructure protection\cite{Buttyan2010} and so on.

Recently, there are increasing numbers of network applications, where
their performance is highly dependent on the high data rate and thus high link capacity requirement.
However, the link
capacity is limited in the WSNs. Thus, many researchers
focus on flow/congestion/rate control designs to achieve efficient and fair resource
allocation in WSNs.
The basic framework of Network Utility Maximization (NUM) proposed in \cite{Kelly1998} has been extended to solve flow control problem  in WSNs \cite{Chen2010}\cite{Hou2008}.
Furthermore, the generalized NUM framework proposed in \cite{Chiang2005} also has been used as a tool of cross-layer design in WSNs\cite{He2010}. However, all these works mentioned above assume that
each link provides a fixed-size transmission ``pipe'' and each user's
utility is a function of transmission rate only.
Furthermore, they don't consider the reliable data delivery requirement,
and implicitly assume an error-free physical layer, which is hard
to achieve in real WSNs.

Since the application performance correlates to
the rates of data obtained reliably in WSNs,
it is vital to guarantee the data delivery reliability
requirement in WSNs.
There are increasing research efforts  to
improve the reliability: reducing the probability of
data loss or error and retransmitting data once loss
or error occurs.
In these work, hop-by-hop recovery\cite{PSFQ2005}, end-to-end recoveryy\cite{Akan2005}, and multi-path forwarding\cite{Deb2003} are the major approaches to achieve the desired reliability.
In PSFQ \cite{PSFQ2005}, the basic premise is to propagate the segments from source nodes in a relatively slow pace and to allow nodes experienced data loss to recover any missing segments from immediate neighbors aggressively.
In ESRT \cite{Akan2005}, exploiting the fact
that the redundancy in sensed data collected in dense WSNs can mitigate channel error
and node failure, the sink adaptively achieves the expected event reliability by controlling the reporting frequency of the source nodes.
In ReInForM \cite{Deb2003}, it  is
proposed to deliver packets at desired reliability by sending multiple copies of each packet along multiple
paths from sources to sink.
Obviously, the data transmission rate and the data delivery reliability are two fundamental, yet conflicting, design objectives in WSNs. There is an intrinsic tradeoff between them. However, all these works mentioned above didn't consider the intrinsic rate-reliability tradeoff problem.
Recent work \cite{Lee2006} has firstly addressed the rate-reliability tradeoff problem explicitly.
Through the extended NUM framework, where the user utility depends on both transmission rate and delivery reliability, the optimal rate-reliability tradeoff can be controlled by adapting channel code rate in each link's physical-layer error correction codes. However, it did not take the energy constraint into consideration, which is one of the most important constraint in WSNs.

Typically, sensor nodes are battery-powered, and battery replacement is impossible in many sensing applications. Energy is a scarce resource, and WSNs have a finite operational lifetime. Hence, network lifetime maximization has been a popular research direction in WSNs, for example, \cite{He2010} has studied the network lifetime maximization problem that jointly considers the physical layer, MAC layer and routing layer. \cite{Liu2010} design an near optimal joint routing-and-sleep-scheduling strategy  to maximize the network lifetime.
\cite{Ehsan2012} propose energy and cross-layer aware routing schemes for multichannel access WSNs that account for radio, MAC contention, and network constraints, aiming to maximize the network lifetime.
\cite{Abdulla2012} propose HYbrid Multi-hop routiNg (HYMN) algorithm, which is a hybrid of the flat multi-hop routing and hierarchical multi-hop routing,  to adequately prolong the lifetime of severely resource-constrained sensor nodes.
\cite{Cheng2009} address joint routing and link rate allocation under bandwidth and energy constraints to prolong network lifetime and to improve throughput.
However, higher data rate leads to greater sensing and communication costs across WSNs, resulting in more energy consumption and shorter network lifetime. Thus, there is an inherent tradeoff between transmission data rate and network lifetime in WSNs. This problem has been extensively studied in recent years\cite{Zhu2007, Nama2006, Chen2009, Zhu2006},
but all these works do not consider the reliability requirement in transmitting the data. On the other hand, to improve the desired reliability,  the approaches, including hop-by-hop recovery\cite{PSFQ2005}, end-to-end recoveryy\cite{Akan2005}, and multi-path forwarding\cite{Deb2003},  generate more data packets to be transmitted,  leading  to more energy expenditure, and shorter network lifetime.
Thus, the network lifetime and the data delivery reliability are also two fundamental, yet conflicting, design objectives in WSNs.

It is clear that there is an inherent tradeoff among the data rate, reliability and network lifetime: A high data rate can be obtained on a link at the expense of lower delivery reliability, which results in more energy consumption and a reduction of network lifetime. Obviously, different applications have very different requirements for rate, reliability, and network lifetime. For example, in emergency rescue and disaster relief, it requires a high data rate and reliability, but does not have high requirements for the network lifetime. However, in precision agriculture, the requirement is to prolong the network lifetime and to improve reliability as much as possible, but the data rate is less demanding.
Although many works have extensively studied the data rate, reliability and network lifetime in recent years separately, so far, no works consider three goals together, and study
the tradeoff among them.
Thus, it is vital to investigate the tradeoff problem of data rate, reliability and network lifetime and to design a efficient distributed algorithm to achieve the optimal rate-reliability-lifetime tradeoff.

In this paper, we address the rate-reliability-lifetime tradeoff problem in energy-constrained WSNs with link capacity constraint, reliability constraint and energy constraint. First, we introduce the weight parameters, which combine the objectives at rate, reliability, and lifetime into a single objective to characterize the tradeoff among them.
However, our new optimization formulation for the rate-reliability-reliability tradeoff is neither separable nor convex. It is difficult to derive a distributed algorithm that converges to the globally optimal solution. Fortunately, through a series of transformations, we convert the formulation into a separable and convex optimization problem.
Then, the Subgradient Dual Decomposition algorithm (SDD) is applied to achieve the optimal solution.
Finally, we investigated the impact of different weight parameter on the rate utility, reliability utility and network lifetime through numerical examples.

 The rest of this paper is organized as follows. Section II describes the system model. Section III  formulate the rate-reliability-lifetime tradeoff problem, and transforms it into a separable and convex optimization problem. Section IV develops the SDD algorithm to solve the tradeoff problem, and proves the convergence of the algorithm. Section V provides numerical examples for the proposed algorithms, and illustrates the optimal rate-reliability-lifetime trade-off.  Finally section VI concludes this paper.

\section{System Model}
In this paper, we consider WSNs consisting of a set of sensor nodes denoted by $\mathcal{S} = \{ 1,2...,S\}$ and a set of sink nodes denoted by $\mathcal{N} = \{ 1,2,...,N\}$. Sensor nodes are battery driven, non-rechargeable and irreplaceable. We assume that the sink nodes have enough energy. The sensor nodes are the sources that collect data and deliver it to any of the sink nodes, possibly over multiple hops. The WSNs is modeled as a connectivity graph $G(\mathcal{V},\mathcal{L})$, where $\mathcal{V} = \mathcal{S} \cup \mathcal{N}$, includes both the sensor nodes and the sink nodes, $\mathcal{L}{\rm{ = \{ 1,2,}}...{\rm{,}}L{\rm{\} }}$ represents the set of logical links between nodes in the network. We assume that the single-path route is adopted in this paper. The key notations used throughout this paper are summarized in Table 1.
\begin{table}[htbp]
 \centering\small
 \begin{threeparttable}
 \caption{\label{tab:table1}Summary of key notations}
  \begin{tabular}{ll}
  \toprule
     Notation   &   Meaning\\
  \midrule
  $\mathcal{S}$       &    The set of sensor nodes\\
  $\mathcal{L}$       &    The set of logical links \\
  $\mathcal{S}{\rm{(}}l{\rm{)}}$ &                     The set of sensor nodes using link $l$\\
  $\mathcal{L}(s)$        &                            The set of links used by sensor node $s$\\
  ${\mathcal{L}_{in}}(s)$  &                           The incoming links set of sensor node $s$\\
  ${\mathcal{L}_{out}}(s)$  &                          The outgoing links set of sensor node $s$\\
  ${\mathcal{S}_{in}}(s)$     &                        The set of source nodes that use sensor node\\
                     &                       $s$  as a relay\\
  ${\mathcal{S}_t}(s)$   &                             The set of sensor nodes that sensor node $s$ \\
                     &                       uses as relays\\
  $l_{in}^{(s,s')}$  &                       The incoming link of sensor node $s$ on the\\
                                    &        path of sensor node $s{\rm{'}}$\\
  $l_{out}^{(s,s')}$                &        The outgoing link of sensor node $s$ on the\\
                            &                path of sensor node $s{\rm{'}}$ \\
  ${l_s}$                        &           The outgoing link that sensor node $s$ uses for\\
                        &                    transmitting its own data\\
  ${C_l}$           &                        The maximum capacity of link $l$\\
  $x_s^{\min }$       &                      The minimum data rates for sensor node $s$\\
  $x_s^{\max }$        &                     The maximum data rates for sensor node $s$\\
  $R_s^{\min }$       &                      The minimum reliability requirement of sensor node $s$\\
  $R_s^{\max }$       &                      The maximum reliability requirement of sensor node $s$\\
  \bottomrule
  \end{tabular}
  \end{threeparttable}
\end{table}
\subsection{Introducing reliability into NUM framework}
Basic NUM framework assumes that each link provides a fixed size transmission `pipe' and each user's utility is only a function of transmission rate. But in many practical systems adapting the physical layer channel coding or transmission diversity, these assumptions will break down. Here, we extended the basic NUM framework to the case that utility of each user depends on both transmission rate and delivery reliability, with an intrinsic tradeoff between them.

Sensor node $s$ sends packets into the encoder of link $l$ at information data rate ${x_s}$. Packets are encoded at the code rate ${r_{l,s}}$, where the code rate ${r_{l,s}}$ is defined by the ratio of the total number of useful information bits to the total number of bits transmitted from the encoder per unit time. The information bits transmitted from the encoder are sent by the wireless link $l$ at the rate ${\phi _{l,s}}$, then ${\phi _{l,s}}$ can be stated as ${\phi _{l,s}} = {{{x_s}} \mathord{\left/
 {\vphantom {{{x_s}} {{r_{l,s}}}}} \right.
 \kern-\nulldelimiterspace} {{r_{l,s}}}}$.

 Since the sum of transmission rates of sensor nodes that traverse the link $l$ can not exceed the maximum link capacity. Thus, we have
 \begin{equation}
 \sum\limits_{s \in \mathcal{S}(l)} {{\phi _{l,s}}}  = \sum\limits_{s \in \mathcal{S}(l)} {\frac{{{x_s}}}{{{r_{l,s}}}} \le {C_l} }\label{eq_2}
\end{equation}
 The error probability of data transmitted by node $s$ using link $l$ is defined as $E({r_{l,s}})$, which is assumed to be an increasing and convex function of $r_{l,s}$.

 Let ${\xi _s}$ denote the end-to-end error probability of each node $s$ , then ${\xi _s}$ is given by
 \[{\xi _s} = 1 - \prod\limits_{l \in \mathcal{L}(s)} {\left( {1 - E({r_{l,s}})} \right)} \]

 In general, the error probability of each link is very small, so the end-to-end error probability of node $s$ can be approximated as ${\xi _s} \approx \sum\nolimits_{l \in \mathcal{L}(s)} {E({r_{l,s}})}$. Let ${R_s}$ denote the reliability of information transmitted by sensor node $s$, then
 \begin{equation}
{R_s} = 1 - {\xi _s} \approx 1 - \sum\nolimits_{l \in \mathcal{L}(s)} {E({r_{l,s}})} \label{eq_3}
\end{equation}

 Now, we introduce the reliability into the NUM framework. We assume that each sensor node $s$ has a utility function ${U_s}\left( {{x_s},{R_s}} \right)$, which is strictly concave increasing functions of the information data rate ${x_s}$ and delivery reliability $R_s$.
 \subsection{Network lifetime maximization problem}
 In a typical sensor network, sensor nodes have much tighter energy constraints than the sink nodes. Hence we will focus only on the energy dissipated in the sensor nodes. Since in most types of sensor nodes, communication modules dominate the energy consumption, we ignore energy consumed by other tasks such as sensing and data processing. So, we adopt the same simple energy consumption model as in \cite{Zhu2006} for the communication module of all nodes. The total power dissipation at node $s$ is given by:
 \begin{equation}
 {\rm{ }}{p_s} = \sum\limits_{l \in {\mathcal{L}_{in}}(s)} {\sum\limits_{s' \in \mathcal{S}(l)} {p_{sl}^r} } {x_{s'}} + \sum\limits_{l \in {\mathcal{L}_{out}}(s)} {\sum\limits_{s' \in \mathcal{S}(l)} {p_{sl}^t} } {x_{s'}}\label{eq_5}
 \end{equation}
 where $p_{sl}^r$ is the energy consumption for receiving unit date from link $l$ at sensor node $s$. In this paper, we assume it to be a constant. $p_{sl}^t$  is the energy consumption for transmitting unit data over link $l$ at sensor node $s$, and is given by:
 \[p_{sl}^t = \psi  + \varsigma d_{sl}^\theta \]
 where $\psi$ is the electronics energy and $\varsigma$ is the amplifier energy, they are constants and depend on the function of the physical layer and the environment factors, ${d_{sl}}$ is the distance of link $l$ at sensor node $s$, $\theta$ is the path loss factor($2 \le \theta  \le 4$).

 We assume that the initial energy of node $s$ is denoted by ${e_s}$, then the lifetime of node $s$ is given by ${T_s} = {{{e_s}} \mathord{\left/
 {\vphantom {{{e_s}} {{p_s}}}} \right.
 \kern-\nulldelimiterspace} {{p_s}}}$. The network lifetime denoted by $T$ is defined as the time at which the first node in the network drains out of energy, then $T = {\min _{s \in \mathcal{S}}}{T_s}$.
\section{Optimal Rate-Reliability-Lifetime Tradeoff}
Now, we formulate the rate-reliability-lifetime tradeoff problem as follows:
\begin{eqnarray}
{\rm{max   }}\;\;\sum\limits_s {\left( {\gamma_s {U_s}({x_s},{R_s}){\rm{ + (1 - }}\gamma_s {\rm{)}}\varpi {{\min }_{{\rm{s}} \in \mathcal{S}}}{T_s}} \right)} \label{eq_7}\\
{\rm{subject}}\;{\rm{to}}\;{\rm{constraints  \;\; (\ref{eq_2}),}}\;(\ref{eq_5})\;,{\rm{and}} \nonumber\\
{R_s} \le 1 - \sum\limits_{l \in \mathcal{L}(s)} {E({r_{l,s}})} {\rm{ ,   }}\;\;s \in \mathcal{S}\label{eq_8}\\
{p_s} = {{{e_s}} \mathord{\left/
 {\vphantom {{{e_s}} {{T_s}}}} \right.
 \kern-\nulldelimiterspace} {{T_s}}}\;,\;\;s \in \mathcal{S}\label{eq_9}\\
x_s^{\min } \le {x_s} \le x_s^{\max }{\rm{ ,   }}\;\;s \in \mathcal{S} \nonumber\\
R_s^{\min } \le {R_s} \le R_s^{\max }{\rm{ ,    }}\;\;s \in \mathcal{S} \nonumber\\
{\rm{0}} \le {r_{l,s}} \le 1{\rm{ ,     }}\;\;l \in \mathcal{L},\;\;s \in \mathcal{S}(l) \nonumber
\end{eqnarray}
$\gamma_s$ ($0\le \gamma_s \le1$) is a weight parameter to combine different objective functions together into a single one. $\varpi$ is a mapping parameter to ensure the objective functions at a same level.
In this problem, constraint (\ref{eq_8}) is reliability constraint for information transmitted by each source. Since the objective function is an increasing function of ${R_s}$, the inequality constraint (\ref{eq_8}) will be satisfied with equality at the optimal solution of ${R_s}$.

It is very difficult to solve the original problem in a distributed manner since the lifetime maximization problem can not be resolved by a distributed manner. We can use a mathematical skill to approximate the network lifetime maximization problem for solving it in a distributed manner. Considering a general utility function ${V^\beta }\left( {\rm{\cdot}} \right)$ defined by
\[{V^\beta }\left( x \right) = \left\{ \begin{array}{l}
\log x,\;\;\;\;\;\;\;\;\;\;\;\beta  = 1\\
\frac{1}{{1 - \beta }}{x^{1 - \beta }},\;\;\;\;\;\beta  > 1
\end{array} \right.\]
Note that maximizing the minimum rate allocation problem for each source  can be approximated by maximizing the aggregate utility when the utility function is given in the above form and $\beta  \to \infty$\cite{Mo2000}. The network lifetime maximization problem(i.e., ${\rm{max}}\;{\min _{{\rm{s}} \in \mathcal{S}}}{T_s}$ ) is similar to max-min rate allocation problem. We introduce a new utility function $V_s^\beta \left( {{T_s}} \right)$ for each sensor node $s$ as a function of its lifetime, which is given by
\[V_s^\beta \left( {{T_s}} \right){\rm{ = }}\frac{1}{{1 - \beta }}{T_s}^{1 - \beta }\]
Then, the maximum network lifetime can be approximated by maximizing the aggregate life time utility, i.e., $\max \;\sum\nolimits_{s \in \mathcal{S}} {V_s^\beta \left( {{T_s}} \right)} $. Since the constraint (\ref{eq_9}) is not convex and separable, we introduce a new variable ${z_s} = 1/{T_s}$, which can be interpreted as the normalized power dissipation of sensor node $s$. Then, the constraint (\ref{eq_9}) becomes
\begin{equation}
 {p_s} = {e_s}{z_s}\;\;\;\forall s \in \mathcal{S}\label{eq_10}
 \end{equation}
As a result, the utility function of network lifetime maximization problem has to change correspondingly:
\begin{equation}
\max \;\sum\nolimits_{s \in \mathcal{S}} {\frac{1}{{1 - \beta }}{z_s}^{\beta  - 1}} \label{eq_11}
\end{equation}
Then, the objective function (\ref{eq_7}) is transformed into
\begin{equation}
W({x_s},{R_s},{z_s}) = \gamma_s {U_s}({x_s},{R_s}) - {\rm{(1}} - \gamma_s {\rm{)}}\frac{\varpi }{{\beta  - 1}}{z_s}^{\beta  - 1}\label{eq_12}
\end{equation}
Obviously, this objective function is strictly concave. Notice that constraint (\ref{eq_2}) in the original problem is not satisfied with the properties of separability and convexity, this leads to the original problem is neither a convex problem nor a separable one. In the next, we will convert the original problem into a separable and convex optimization problem through a series of transformations.

First, we introduce a auxiliary variables ${c_{l,s}}$, which can be interpreted as the allocated transmission capacity to sensor node $s$ on the link $l$. Then, the constraint (\ref{eq_2}) is decomposed into two constraints i.e.,
\begin{equation}
\frac{{{x_s}}}{{{r_{l,s}}}} \le {c_{l,s}} ,  l \in \mathcal{L},s \in \mathcal{S}(l)\label{eq_13}
\end{equation}
\begin{equation}
\sum\nolimits_{s \in \mathcal{S}(l)} {{c_{l,s}}}  \le {C_l},{\rm{ }}l \in \mathcal{L}\label{eq_14}
\end{equation}

Notice that the inequality constraint (\ref{eq_13}) is still inseparable. We take logarithm on both sides of this constraint, i.e., $\log {x_s} - \log {r_{l,s}}   \le  \log {c_{l,s}}$. Let ${x'_s} = \log {x_s}$ (i.e., ${x_s} = {e^{{x'_s}}}$), then the constraint (\ref{eq_13}) is changed into
\begin{equation}
{x'_s} - \log {r_{l,s}}   \le  \log {c_{l,s}} ,\;\;l \in \mathcal{L},s \in \mathcal{S}(l)\label{eq_15}
\end{equation}
Correspondingly, the objective function is transformed into
\[W'({x'_s},{R_s},{z_s}) = \gamma_s {U'_s}({x'_s},{R_s}) - {\rm{(1}} - \gamma_s {\rm{)}}\frac{\varpi }{{\beta  - 1}}{z_s}^{\beta  - 1}\]
where ${U'_s}({x'_s},{R_s}) = {U_s}({e^{{x'_s}}},{R_s})$. However, ${U'_s}({x'_s},{R_s})$ may not be a concave function , even though ${U_s}({x_s},{R_s})$ is a concave function. The lemma 2 shown in \cite{Lee2006} provides a sufficient condition for its concavity, under which, the objective function $W'({x'_s},{R_s},{z_s})$  is also a concave function.

\begin{prop}
The rate-reliability-lifetime tradeoff problem is equivalent to the convex problem
\begin{equation}\label{eq_17}
\begin{array}{l}
\;{\rm{max   }}\;\;\sum\limits_s {W'({x'_s},{R_s},{z_s})} \\
\;\;\;\;\;\;\;\;\;{\rm{subject}}\;{\rm{to}}\;{\rm{constraints \;\; (\ref{eq_8}),}}\;{\rm{(\ref{eq_14}), (\ref{eq_15}) \;\;and}}\\
\sum\limits_{\scriptscriptstyle l \in {\mathcal{L}_{in}}(s)} {\sum\limits_{\scriptscriptstyle s' \in \mathcal{S}(l)} {p_{sl}^r} } {e^{{x'_{s'}}}} + \sum\limits_{\scriptscriptstyle l \in {\mathcal{L}_{out}}(s)} {\sum\limits_{\scriptscriptstyle s' \in \mathcal{S}(l)} {p_{sl}^t} } {e^{{x'_{s'}}}} \le {e_s}{z_s}\;{\rm{ ,   }}s \in \mathcal{S}\\
{\rm{ \hspace{15mm}}}\;\;{x'_s}^{\min } \le {x'_s} \le {x'_s}^{\max }{\rm{ ,   }}\;\;s \in \mathcal{S}\\
{\rm{\hspace{17mm}}}R_s^{\min } \le {R_s} \le R_s^{\max }{\rm{ ,    }}\;\;s \in \mathcal{S}\\
{\rm{ \hspace{15mm}}}\;{\rm{ }}\;{\rm{0}} \le {r_{l,s}} \le 1\;{\rm{,    }}\;\;l \in \mathcal{L},\;\;s \in \mathcal{S}(l)\\
\;\;\;\;\;\;\;\;\;\;\;\;\;\;\;0\; \le \;{c_{l,s}} \le {C_l}\;,\;\;\;l \in \mathcal{L},\;\;s \in \mathcal{S}(l)
\end{array}
\end{equation}
where ${x'_s}^{\min } = \log x_s^{\min }$ , and ${x'_s}^{\max }{\rm{ = }}\log x_s^{\max }$.
\end{prop}

\textbf{Proof}
First, the constraints (\ref{eq_5}) and (\ref{eq_10}) can be combined into a single one
$${\rm{ }}{e_sz_s} = \sum\limits_{l \in {\mathcal{L}_{in}}(s)} {\sum\limits_{s' \in \mathcal{S}(l)} {p_{sl}^r} } {x_{s'}} + \sum\limits_{l \in {\mathcal{L}_{out}}(s)} {\sum\limits_{s' \in \mathcal{S}(l)} {p_{sl}^t} } {x_{s'}}$$
By variable substitution ${x_s} = {e^{{x'_s}}}$, the above equality constraint reduces to
\begin{equation}
\sum\limits_{l \in {\mathcal{L}_{in}}(s)} {\sum\limits_{s' \in \mathcal{S}(l)} {p_{sl}^r} } {e^{{x'_{s'}}}} + \sum\limits_{l \in {\mathcal{L}_{out}}(s)} {\sum\limits_{s' \in \mathcal{S}(l)} {p_{sl}^t} } {e^{{x'_{s'}}}}{\rm{ = }}{e_s}{z_s}\label{eq_16}
\end{equation}
Hence, the rate-reliability-lifetime tradeoff problem can be expressed as
\begin{equation}\label{eq_17_eq}
\begin{array}{l}
\;{\rm{max   }}\;\;\sum\limits_s {W'({x'_s},{R_s},{z_s})} \\
\;\;\;\;\;\;\;\;\;{\rm{subject}}\;{\rm{to}}\;{\rm{constraints \;\; (\ref{eq_8}),}}\;{\rm{(\ref{eq_14}), (\ref{eq_15}) \;\;and}}\\
\sum\limits_{\scriptscriptstyle l \in {\mathcal{L}_{in}}(s)} {\sum\limits_{\scriptscriptstyle s' \in \mathcal{S}(l)} {p_{sl}^r} } {e^{{x'_{s'}}}} + \sum\limits_{\scriptscriptstyle l \in {\mathcal{L}_{out}}(s)} {\sum\limits_{\scriptscriptstyle s' \in \mathcal{S}(l)} {p_{sl}^t} } {e^{{x'_{s'}}}} ={e_s}{z_s}\;{\rm{ ,   }}s \in \mathcal{S}\\
{\rm{ \hspace{15mm}}}\;\;{x'_s}^{\min } \le {x'_s} \le {x'_s}^{\max }{\rm{ ,   }}\;\;s \in \mathcal{S}\\
{\rm{\hspace{17mm}}}R_s^{\min } \le {R_s} \le R_s^{\max }{\rm{ ,    }}\;\;s \in \mathcal{S}\\
{\rm{ \hspace{17mm}}}\;{\rm{ }}\;{\rm{0}} \le {r_{l,s}} \le 1\;{\rm{,    }}\;\;l \in \mathcal{L},\;\;s \in \mathcal{S}(l)\\
\;\;\;\;\;\;\;\;\;\;\;\;\;\;\;\;\;\;\;0\; \le \;{c_{l,s}} \le {C_l}\;,\;\;\;l \in \mathcal{L},\;\;s \in \mathcal{S}(l)
\end{array}
\end{equation}
Due to the objective of network lifetime maximization (equivalently, minimizing $z_s$'s), it is easily known that, at the optimality of the problem \eqref{eq_17}, the inequality constraint
$$\sum\limits_{l \in {\mathcal{L}_{in}}(s)} {\sum\limits_{s' \in \mathcal{S}(l)} {p_{sl}^r} } {e^{{x'_{s'}}}} + \sum\limits_{l \in {\mathcal{L}_{out}}(s)} {\sum\limits_{s' \in \mathcal{S}(l)} {p_{sl}^t} } {e^{{x'_{s'}}}}{\rm{ \le }}{e_s}{z_s}$$ must hold with equality. Hence, the problem \eqref{eq_17_eq} can be equivalent to \eqref{eq_17} which is a convex problem. This completes the proof.

\section{Subgradient Dual Decomposition}
In this section, we will use subgradient dual decomposition approach to solve the problem (\ref{eq_17}). We write down the Lagrangian function associated with the problem (\ref{eq_17}) as in the first equality of (\ref{eq_18}) and rearrange the Lagrangian function as in the last equality of Eq. (\ref{eq_18}), where the new sets ${\mathcal{S}_{in}}(s)$, ${\mathcal{S}_{t}}(s)$ and variables $l_{in}^{(s,s')}$, $l_{out}^{(s,s')}$, ${l_s}$ are introduced for ease of separation, ${\lambda _{l,s}}$, ${\mu _s} $  and ${\nu _s}$  are the Lagrange multipliers which can be respectively interpreted as congestion price on link $l$, reliability price and energy consumption price on sensor node $s$.
\begin{figure*}[!t]
\begin{equation}
\begin{array}{l}
\mathcal{L}\left( {{\bf{x'}},{\bf{R}},{\bf{r}},{\bf{c}},{\bf{z}},{\bm{\lambda }},{\bm{\mu }},{\bm{\nu }}} \right)\\
\mathop  = \sum\limits_s {W'({x'_s},{R_s},{z_s})}  + \sum\limits_l {\sum\limits_{s \in \mathcal{S}\left( l \right)} {{\lambda _{l,s}}\left( {\log {c_{l,s}} + \log {r_{l,s}} - {x'_s}} \right)} }  + \sum\limits_s {{\mu _s}\left( {1 - \sum\limits_{l \in \mathcal{L}(s)} {E({r_{l,s}})} {\rm{ }} - {R_s}} \right)} \\
\;\;\;\;{\rm{            }}\; + \sum\limits_s {{\nu _s}\left( {{e_s}{z_s} - \sum\limits_{l \in {\mathcal{L}_{in}}(s)} {\sum\limits_{s' \in \mathcal{S}(l)} {p_{sl}^r{e^{{x'_{s'}}}}} }  - \sum\limits_{l \in {\mathcal{L}_{out}}(s)} {\sum\limits_{s' \in \mathcal{S}(l)} {p_{sl}^t{e^{{x'_{s'}}}}} } } \right)} \\
\mathop  = \sum\limits_s {\left\{ {W'({x'_s},{R_s},{z_s}) - {x'_s}\sum\limits_{{\rm{ }}l \in \mathcal{L}\left( s \right)} {{\lambda _{l,s}}}  - {\mu _s}{R_s} + {\nu _s}{e_s}{z_s} - \sum\limits_{s' \in {\mathcal{S}_{in}}(s)} {{\nu _s}p_{s,l_{in}^{(s,s')}}^r{e^{{x'_{s'}}}}} \; - \sum\limits_{s' \in {\mathcal{S}_{in}}(s)} {{\nu _s}p_{s,l_{out}^{(s,s')}}^t{e^{{x'_{s'}}}}}  - {\nu _s}{e^{{x'_s}}}p_{s{l_s}}^t} \right\}} \\
\;\;\;\;\;\;{\rm{ + }}\sum\limits_l {\left\{ {\sum\limits_{s \in \mathcal{S}\left( l \right)} {({\lambda _{l,s}}(\log {c_{l,s}} + \log {r_{l,s}}) - {\mu _s}E({r_{l,s}}))} } \right\}}  + \sum\limits_s {{\mu _s}} {\rm{   }}\\
\mathop  = \sum\limits_s {\left\{ {W'({x'_s},{R_s},{z_s}) - {\lambda ^s}{x'_s} - {\mu _s}{R_s} + {\nu _s}{e_s}{z_s} - {e^{{x'_s}}}\sum\limits_{s' \in {\mathcal{S}_t}(s)} {{\nu _{s'}}p_{s',l_{in}^{(s',s)}}^r} \;\; - {e^{{x'_s}}}\sum\limits_{s' \in {\mathcal{S}_t}(s)} {{\nu _{s'}}p_{s',l_{out}^{(s',s)}}^t}  - {\nu _s}{e^{{x'_s}}}p_{s{l_s}}^t} \right\}} \\
\;\;\;\;\;\;{\rm{ + }}\sum\limits_l {\left\{ {\sum\limits_{s \in \mathcal{S}\left( l \right)} {({\lambda _{l,s}}(\log {c_{l,s}} + \log {r_{l,s}}) - {\mu _s}E({r_{l,s}}))} } \right\}}  + \sum\limits_s {{\mu _s}} {\rm{        }}\\
\mathop  = \sum\limits_s {\left\{ {W'({x'_s},{R_s},{z_s}) - {\lambda ^s}{x'_s} - {\mu _s}{R_s} + {\nu _s}{e_s}{z_s} - {e^{{x'_s}}}\sum\limits_{s' \in {\mathcal{S}_t}(s)} {{\nu _{s'}}\left( {p_{s',l_{in}^{(s',s)}}^r + p_{s',l_{out}^{(s',s)}}^t} \right)}  - {\nu _s}{e^{{x'_s}}}p_{s{l_s}}^t} \right\}} \;\\
\;\;\;\;\;\;{\rm{ + }}\sum\limits_l {\left\{ {\sum\limits_{s \in \mathcal{S}\left( l \right)} {({\lambda _{l,s}}(\log {c_{l,s}} + \log {r_{l,s}}) - {\mu _s}E({r_{l,s}}))} } \right\}}  + \sum\limits_s {{\mu _s}}\\
\mathop  = \sum\limits_s {\left\{ {W'({x'_s},{R_s},{z_s}) - {\lambda ^s}{x'_s} - {\mu _s}{R_s} + {\nu _s}{e_s}{z_s} - {e^{{x'_s}}}\sum\limits_{s' \in {\mathcal{S}_t}(s)} {{\nu _{s'}}{p^{(s',s)}}}  - {\nu _s}{e^{{x'_s}}}p_{s{l_s}}^t} \right\}} \;\\
\;\;\;\;\;\;{\rm{ + }}\sum\limits_l {\left\{ {\sum\limits_{s \in \mathcal{S}\left( l \right)} {({\lambda _{l,s}}(\log {c_{l,s}} + \log {r_{l,s}}) - {\mu _s}E({r_{l,s}}))} } \right\}}  + \sum\limits_s {{\mu _s}} \label{eq_18}
\end{array}
\end{equation}
\end{figure*}

In Eq. (\ref{eq_18}), $\mathop \lambda \nolimits^s {\rm{ = }}\sum\nolimits_{{\rm{ }}l \in \mathcal{L}\left( s \right)} {{\lambda _{l,s}}}$ , i.e., the end-to-end congestion price at the sensor node $s$ , and ${p^{(s',s)}} = p_{s',l_{in}^{(s',s)}}^r + p_{s',l_{out}^{(s',s)}}^t$ , i.e., the power dissipation of relaying unit data from node $s$ at node $s'$ ,$s' \in {\mathcal{S}_t}(s)$ , $s \in \mathcal{S}$ . The Lagrange dual function is given by
\begin{equation}
\begin{array}{l}
G\left( {{\mbox{\boldmath$\lambda$},\mbox{\boldmath$\mu$},\mbox{\boldmath$\nu$}}} \right) = {\rm{max \;\;\;\;}}\mathcal{L}\left( {{{\bf{x}}^{\bf{'}}},{\bf{R}},{\bf{r}},{\bf{c}},{\bf{z}},{\mbox{\boldmath$\lambda$},\mbox{\boldmath$\mu$},\mbox{\boldmath$\nu$}}} \right)\\
{\rm{                }}\;\;\;\;\;\;\;\;\;\;\;\;\;\;\;\;\;\;\;\;\;\;{\rm{ subject}}\;{\rm{to \;\;\;\;   }}{{\bf{x}}^{{\bf{'min}}}}\preceq{{\bf{x}}^{\bf{'}}}\preceq{{\bf{x}}^{{\bf{'max}}}}\\
{\bf{         }}\;\;\;\;\;\;\;\;\;\;\;{\bf{ }}\;\;\;\;\;\;\;\;\;\;\;\;\;\;\;\;\;\;\;\;\;\;\;\;\;\;\;\;\;\;\;{{\bf{R}}^{{\bf{min}}}}\preceq{\bf{R}}\preceq{{\bf{R}}^{{\bf{max}}}}{\bf{ }}\\
{\bf{         }}\;\;\;\;\;\;\;\;\;\;\;\;\;\;\;\;\;\;\;\;\;\;\;\;\;\;\;\;\;\;\;\;\;\;\;\;\;\;\;\;\;\;\;{\bf{ 0}}\preceq{\bf{r}}\preceq{\bf{1 }}\\
\;\;\;\;\;\;\;\;\;\;\;\;\;\;\;\;\;\;\;\;\;\;\;\;\;\;\;\;\;\;\;\;\;\;\;\;\;\;\;\;\;\;\;{\bf{c}} \in \mathcal{C} \label{eq_19}
\end{array}
\end{equation}
where $\mathcal{C} = \{ ({c_{l,s}})l \in \mathcal{L},s \in \mathcal{S}(l)|\sum\nolimits_{s \in \mathcal{S}(l)} {{c_{l,s}}}  \le {C_l},\;l \in \mathcal{L},0\; \le \;{c_{l,s}} \le {C_l},\;l \in \mathcal{L},s \in \mathcal{S}(l)\}$.

The dual problem corresponding to problem (\ref{eq_17}) is then given by
\begin{equation}
\mathop {{\rm{min }}}\limits_{{\bm{\lambda }}\succeq0,{\bm{\mu }}\succeq0,{\bm{\nu }}\succeq0} {\rm{  }}G\left( {{\bm{\lambda ,\mu ,\nu }}} \right){\rm{ }}\label{eq_20}
\end{equation}

The key of the dual-decomposition algorithm is solving the problem (\ref{eq_19}) separately and distributively. With the separation in Eq. (\ref{eq_18}), maximization the Lagrangian over $({{\bf{x}}^{\bf{'}}},{\bf{R}},{\bf{r}},{\bf{c}},{\bf{z}})$ can be done in parallel at each sensor node $s$
\begin{equation}
\begin{array}{l}
{\rm{ max    }\hspace{4mm}}W'({x'_s},{R_s},{z_s}) - {\lambda ^s}{x'_s} - {\mu _s}{R_s}\\
\;\;\;\;\;\;\;\;\;\; + {\nu _s}{e_s}{z_s} - {e^{{x'_s}}}\sum\limits_{\mathclap{s' \in {\mathcal{S}_t}(s)}} {{\nu _{s'}}{p^{(s',s)}}}  - {\nu _s}{e^{{x'_s}}}p_{s{l_s}}^t\\
\;{\rm{  }}s.t.{\rm{  \hspace{4mm}     }}x_s^{'\min } \le {{x'}_s} \le x_s^{'\max }{\rm{ }}\\
\;{\rm{    \hspace{9mm}         }}R_s^{\min } \le {R_s} \le R_s^{\max }\label{eq_21}
\end{array}
\end{equation}
and at each link $l$
\begin{equation}
\begin{array}{l}
\max \;\;\;\;\;\sum\limits_{s \in \mathcal{S}\left( l \right)} {({\lambda _{l,s}}} (\log {c_{l,s}} + \log {r_{l,s}}) - {\mu _s}E({r_{l,s}}))\\
\;s.t.\;\;\;\;\;\;\;\sum\limits_{s \in \mathcal{S}(l)} {{c_{l,s}} \le {C_l} } \\
\;\;\;\;\;\;\;\;\;\;\;\;\;\;\;0\; \le \;{c_{l,s}} \le {C_l}\;,\;\;\;\;s \in \mathcal{S}(l)\\
\;\;\;\;\;\;\;\;\;\;\;\;\;\;\;{\rm{0}} \le {r_{l,s}} \le 1{\rm{ ,          }}\;\;s \in \mathcal{S}(l)\label{eq_22}
\end{array}
\end{equation}
The problem (\ref{eq_22}) can be further decomposed into two sub-problems as follows:

Link-layer sub-problem
\begin{equation}
\begin{array}{l}
\max \;\;\;\;\;\sum\limits_{s \in \mathcal{S}\left( l \right)} {{\lambda _{l,s}}} \log {c_{l,s}}\\
\;s.t.\;\;\;\;\;\;\;\sum\limits_{s \in \mathcal{S}(l)} {{c_{l,s}} \le {C_l} } \\
\;\;\;\;\;\;\;\;\;\;\;\;\;\;\;0\; \le \;{c_{l,s}} \le {C_l}\;,\;\;\;\;s \in \mathcal{S}(l)\label{eq_23}
\end{array}
\end{equation}
and physical-layer sub-problem for sensor node $s$,\;$s \in \mathcal{S}(l)$
\begin{equation}
\begin{array}{l}
\max \;\;\;\;\;{\lambda _{l,s}}\log {r_{l,s}} - {\mu _s}E({r_{l,s}})\\
\;s.t.\;\;\;\;\;\;\;\;{\rm{0}} \le {r_{l,s}} \le 1{\rm{ ,          }}s \in \mathcal{S}(l)\label{eq_24}
\end{array}
\end{equation}
Once the problem (\ref{eq_19}) is solved, the subgradients of the dual function with respect to the dual variables can be calculated easily and the dual variables for solving the dual problem (\ref{eq_20}) can be iteratively updated by using subgradient projection method \cite{Bertsekas1999} as follows:\\
Congestion price update at each link $l$, $l \in \mathcal{L}$, $s \in \mathcal{S}(l)$
\begin{equation}
\begin{array}{l}
{\lambda _{l,s}}\left( {t + 1} \right)\\
\;\;\;\; = {\left[ {{\lambda _{l,s}}\left( t \right) - \delta \left( t \right)\left( {\log {c_{l,s}}(t) + \log {r_{l,s}}(t) - {{x'}_s}(t)} \right)} \right]^{\rm{ + }}}\;\;\;\\
\;\;\;\;{\rm{ = }}{\left[ {{\lambda _{l,s}}\left( t \right) - \delta \left( t \right)\left( {\log {c_{l,s}}(t) + \log {r_{l,s}}(t) - \log {x_s}(t)} \right)} \right]^{\rm{ + }}}\label{eq_25}
\end{array}
\end{equation}
Reliability price update at each sensor node $s$, $s \in \mathcal{S}$
\begin{equation}
\begin{array}{l}
{\mu _s}\left( {t + 1} \right)\\
\;\;\;\;\; = \mathop {\left[ {{\mu _s}\left( t \right) - \zeta \left( t \right)\left( {1 - \sum\limits_{l \in \mathcal{L}(s)} {E({r_{l,s}}(t))} {\rm{ }} - {R_s}\left( t \right)} \right)} \right]}\nolimits^ +  \\
\;\;\;\;\;\mathop { = \left[ {{\mu _s}\left( t \right) - \zeta \left( t \right)\left( {\mathop R\nolimits^s \left( t \right) - {R_s}\left( t \right)} \right)} \right]}\nolimits^ +  \label{eq_26}
\end{array}
\end{equation}
where $\mathop R\nolimits^s (t) = 1 - \sum\limits_{l \in \mathcal{L}(s)} {E({r_{l,s}}(t))}$ with an interpretation of end-to-end reliability at a sensor node $s$.
\begin{figure*}[!t]
\begin{equation}
\begin{array}{l}
{\nu _s}\left( {t + 1} \right)\\
\;\;\;\; = {\left[ {{\nu _s}\left( t \right) - \vartheta \left( t \right)\left( {{e_s}{z_s}(t) - \sum\limits_{l \in {\mathcal{L}_{in}}(s)} {\sum\limits_{s' \in \mathcal{S}(l)} {p_{sl}^r{e^{{x'_{s'}}(t)}}} }  - \sum\limits_{l \in {\mathcal{L}_{out}}(s)} {\sum\limits_{s' \in \mathcal{S}(l)} {p_{sl}^t{e^{{x'_{s'}}(t)}}} } } \right)} \right]^ + }\;\\
\;\;\;\; = {\left[ {{\nu _s}\left( t \right) - \vartheta \left( t \right)\left( {{e_s}{z_s}(t) - \sum\limits_{s' \in {\mathcal{S}_{in}}(s)} {{e^{{x'_{s'}}(t)}}\left( {p_{sl_{in}^{(s,s')}}^r + p_{sl_{out}^{(s,s')}}^t} \right)}  - \;{e^{{x'_s}(t)}}p_{s{l_s}}^t} \right)} \right]^ + }\\
\;\;\;\; = {\left[ {{\nu _s}(t) - \vartheta (t)\left( {{e_s}{z_s}(t) - \sum\limits_{s' \in {\mathcal{S}_{in}}(s)} {{x_{s'}}(t){p^{(s,s')}}}  - \;{x_s}(t)p_{s{l_s}}^t} \right)} \right]^ + }\label{eq_27}
\end{array}
\end{equation}
\end{figure*}\\\\
The energy consumption price ${\nu_s}$ is updated according to (\ref{eq_27}) (see the top of the next page), $s \in \mathcal{S}$, where ${p^{(s,s')}} = p_{s,l_{in}^{(s,s')}}^r + p_{s,l_{out}^{(s,s')}}^t$ , i.e., the power dissipation of node $s$ for relaying unit data from node $s'$  , $s' \in {\mathcal{S}_{in}}(s)$ , $s \in \mathcal{S}$ . In the above formulas, ${\left[ w \right]^{\rm{ + }}} = \max \{ 0,w\} $ , and $\delta \left( t \right)$ , $\zeta \left( t \right)$ and $\vartheta \left( t \right)$ are positive scalar step size.

We will summarize the distributed algorithm for rate-reliability-lifetime tradeoff as follows, where each sensor node and each link will solve their own problems with only local information. The information exchange between sensor node and link in the distributed algorithm SDD is given in Fig. 1.

\noindent\rule{500pt}{1pt}
\textbf{SDD: Subgradient Dual Decomposition Algorithm}\\
\noindent\rule{500pt}{0.65pt}
\\at each iteration $t$
  \\\indent \textbf{at each sensor node $s$}
  \begin{enumerate}
  \item Rate, reliability and lifetime update:
  \begin{itemize}
    \item Node $s$ receives the link congestion price ${\lambda _{l,s}}(t)$ of link $l$ from the network, $l \in \mathcal{L}(s)$, and calculates $\mathop \lambda \nolimits^s {\rm{(t)}}$ according to $\mathop \lambda \nolimits^s {\rm{(t) = }}\sum\nolimits_{{\rm{ }}l \in \mathcal{L}\left( s \right)} {{\lambda _{l,s}}(t)}$.
    \item Node $s$ receives the energy consumption price ${\nu_{s'}}(t)$ of sensor nodes $s'$ who relay data packets for node $s$, $s' \in {\mathcal{S}_t}(s)$.
    \item Node $s$ locally solves the problem (\ref{eq_21}) for the given ${\mu _s}(t)$, ${\nu _s}(t)$, ${\nu_{s'}}(t)$ and $\mathop \lambda \nolimits^s {\rm{(t)}}$. Update the information rate ${x_s}\left( {t + 1} \right)$(where ${x_s}\left( {t + 1} \right) = {e^{{x'_s}(t + 1)}}$), information reliability ${R_s}(t + 1)$ and node lifetime ${T_s}(t + 1)$(where ${T_s}(t + 1) = {1 \mathord{\left/
 {\vphantom {1 {{z_s}\left( {t + 1} \right)}}} \right.
 \kern-\nulldelimiterspace} {{z_s}\left( {t + 1} \right)}}$ ).
    \item Broadcasts the new rate ${x_s}\left( {t + 1} \right)$ to links that sensor node $s$ uses.
\end{itemize}
  \item Reliability price update:
  \begin{itemize}
    \item Node $s$ receives code rate ${r_{l,s}}(t)$ of link $l$, $l \in \mathcal{L}(s)$, and computes the end-to-end reliability for given $\mathop R\nolimits^s (t) = 1 - \sum\nolimits_{l \in \mathcal{L}(s)} {E({r_{l,s}}(t))}$, then, updates its reliability price according to Eq. (\ref{eq_26}).
    \item Broadcasts the new reliability price ${\mu _s}\left( {t + 1} \right)$ to the links that sensor node $s$ uses.
    \end{itemize}
  \item Energy consumption price update:
  \begin{itemize}
    \item Node $s$ receives ${x_{s'}}(t)$ of the sensor nodes that use node $s$ relaying their data packets, $s' \in {\mathcal{S}_{in}}(s){\rm{ }}$, and updates its energy consumption price according to Eq. (\ref{eq_27}).
    \item Broadcasts the new energy consumption price ${\nu _s}\left( {t + 1} \right)$ to the nodes that use node $s$ relaying their data packets.
   \end{itemize}
  \end{enumerate}
  \indent\indent \textbf{at each link $l$}:
  \begin{enumerate}
    \item Auxiliary variables ${c_{l,s}}$ update:
    \\Link $l$ update ${c_{l,s}}(t)$ by locally solving the link-layer problem (\ref{eq_23}) for given ${\lambda _{l,s}}(t)$.
    \item Code rate ${r_{l,s}}$ update:
    \begin{itemize}
    \item Link $l$ receives reliability price ${\mu _s}(t)$ of sensor node $s$, $s \in \mathcal{S}(l)$ , then, update ${r_{l,s}}\left( {t + 1} \right)$  by locally solving the physical-layer problem (\ref{eq_24}) for given ${\mu _s}(t)$ and ${\lambda _{l,s}}(t)$.
    \item Broadcasts new code rate ${r_{l,s}}\left( {t + 1} \right)$  to the sensor nodes that use link $l$, $s \in \mathcal{S}(l)$.
    \end{itemize}
    \item Congestion price update:
    \begin{itemize}
    \item Link $l$ receives data at rate ${x_s}\left( t \right)$ from the sensor nodes that use link $l$, $s \in \mathcal{S}(l)$ , and updates its congestion price according to Eq. (\ref{eq_25}) .
    \item Broadcasts the new congestion price ${\lambda _{l,s}}\left( {t + 1} \right)$  to the nodes that use link $l$, $s \in \mathcal{S}(l)$.
    \end{itemize}
\end{enumerate}
\noindent\rule{500pt}{1pt}\\

 \cite{Shor1985} shows the global convergence of the proposed dual-decomposition based algorithm. For convenience,we state the convergence result in Theorem 1.

  \textbf{\emph{Theorem 1:}} By the above distributed algorithm, dual variables (\mbox{\boldmath$\lambda$}\emph{(t)}, \mbox{\boldmath$\mu$}\emph{(t)}, \mbox{\boldmath$\nu$}\emph{(t)}) converge to the optimal dual solutions (${{\bm{\lambda }}^{\bm{*}}}{\bm{,}}\;{{\bm{\mu }}^{\bm{*}}}{\bm{,}}\;{{\bm{\nu }}^{\bm{*}}}$ ), if the setpsizes are chosen such that $\delta \left( t \right) \to 0,\;\sum\limits_{t = 1}^\infty  {\delta \left( t \right) = \infty }$ , $\zeta \left( t \right) \to 0\;\sum\limits_{t = 1}^\infty  {\zeta \left( t \right)}  = \infty$  and $\vartheta \left( t \right) \to 0,\;\sum\limits_{t = 1}^\infty  {\vartheta \left( t \right) = \infty }$.

 On the other hand, since the tradeoff problem (\ref{eq_17}) is a convex optimization problem, by Lagrange-duality theory\cite{Bertsekas1999}, we conclude that, the corresponding primal variables (${{\bf{x}}^{\bf{*}}}{\bf{,}}\;{{\bf{R}}^{\bf{*}}}{\bf{,}}\;{{\bf{z}}^{\bf{*}}}{\bf{,}}\;{{\bf{c}}^{\bf{*}}}{\bf{,}}\;{{\bf{r}}^{\bf{*}}}$ ) are the optimal solutions of the problem (\ref{eq_17}).

\section{Numerical Examples}
 In this section, we present numerical examples for the proposed algorithm by considering a sensor network, shown in Fig. \ref{fig:fig2}. In this network, we have six sensor nodes indexed 1-6, and one sink node. They communicate through seven links. The locations of sensor nodes are randomly generated over a 100m $ \times $ 100m square area.

We set the objective function $W({x_s},{R_s},{z_s})$  in the following form
\[\begin{array}{l}
W({x_s},{R_s},{z_s}) = \gamma_s \varphi_s \frac{{x_s^{1 - \alpha } - x_s^{\min (1 - \alpha )}}}{{x_s^{\max (1 - \alpha )} - x_s^{\min (1 - \alpha )}}}\\
\;\;\;\;\;\;\;\;\;\;\;\;\;\;\;\;\;\;\;\;\; + \gamma_s (1 - \varphi_s )\frac{{R_s^{1 - \alpha } - R_s^{\min (1 - \alpha )}}}{{R_s^{\max (1 - \alpha )} - R_s^{\min (1 - \alpha )}}}\;\\
\;\;\;\;\;\;\;\;\;\;\;\;\;\;\;\;\;\;\;\;\; - {\rm{(1}} - \gamma_s {\rm{)}}\frac{\varpi }{{\beta  - 1}}{z_s}^{\beta  - 1}\;\;\;\;\alpha  > 0,\;\alpha  \ne 1
\end{array}\]
where $\varphi_s $ ($0 \le \varphi_s  \le 1$ ) is the weight parameter that weights the relative importance of information data rate and reliability. The three parts of the expression of the objective function represent the rate utility, reliability utility and network lifetime, respectively.
$E({r_{l,s}})$ is assumed to be
of the following form:
 \[E\left( {{r_{l,s}}} \right){\rm{ = }}\frac{1}{2}\exp \left( { - \kappa \left( {1 - {r_{l,s}}} \right)} \right)\]
 where $\kappa $ is the code block length used by the encoder.
In our experiments, the constant parameters are set as follows: $x_s^{\min } = 0.1\;Mbps$, $x_s^{\max } = 2.0\;Mbps$, $R_s^{\min } = 0.9$, $R_s^{\max } = 1$, $\alpha {\rm{ = }}1.1$, $\beta {\rm{ = }}9$ and the mapping parameter $\varpi {\rm{ = }}3.2768 \times {10^{32}}$.

 We assume that there is a routing mechanism in place to find a route for each sensor node. The routes of each flow have been drawn in Fig. \ref{fig:fig2}. The capacity of links $a-g$ are set to be ${C_l} = [2; 3; 2.5; 3; 4; 2.5; 4]\;(Mbps)$.  The initial energy of sensor nodes are set to be ${\bm{e}} = {\rm{[3000;}}\;{\rm{2800;}}\;{\rm{2500;}}\;{\rm{2200;}}\;{\rm{2600;}}\;{\rm{2000]}}\;{\rm{(J)}}$. The power dissipation at node $s$  is determined by (\ref{eq_5}) in section II, where $\psi  = {{50nJ} \mathord{\left/
 {\vphantom {{50nJ} b}} \right.
 \kern-\nulldelimiterspace} b}$ , $\varsigma  = 0.0013pJ/b/{m^4}$ , $\theta {\rm{ = }}4$  and $p_{sl}^r{\rm{ = }}50nJ/b$ \cite{Heinzelman2002}.

 \subsection{Convergence performance of SDD}
 In the section, we depict the convergence performance of SDD algorithm. We only show the convergence results of nodes rates, the convergence results of others are omitted due to space limitations. Both the weight parameter $\gamma_s $  and $\varphi_s $ are set to 0.8, and the rates of sensor nodes are shown in Fig. \ref{fig:fig3}. Notice that the rates of all sensor nodes change sharply at the beginning of the iteration, and then converge to the optimal solution after about 200 iterations.

Fig.\ref{fig:fig4} depicts the convergent behavior of total utility solved by SDD. The blue dotted line denotes the optimal value of the total utility solved by the centralized algorithm. Obviously, the values of the total utility solved by SDD converge to the optimal value after about 200 iterations.

Then, both the weight parameter $\gamma_s $  and $\varphi_s $ are increased to 0.97, in which the reliability and network lifetime is almost out of consideration. The convergent results of nodes rates and total utility of SDD are shown in Fig. \ref{fig:fig5} and Fig. \ref{fig:fig6}, respectively. Notice that the optimal rates of all the sensor nodes are larger than those in Fig. \ref{fig:fig3}, since we increase $\gamma_s $  and $\varphi_s $  to obtain more rate utility while reducing the data reliability and network lifetime.

Furthermore, both the weight parameter $\gamma_s $  and $\varphi_s $ are set to 0.5 , in which the network lifetime is more consideration than the rate and reliability. The convergent results of nodes rates of SDD are shown in Fig. \ref{fig:fig7}. Compare with Fig. \ref{fig:fig3} and Fig. \ref{fig:fig5}, the optimal rates of all the sensor nodes are very small. In this case, the sensor nodes can save more energy to extend the network lifetime. Fig.\ref{fig:fig8} depicts the convergent behavior of total utility. Compare with Fig. \ref{fig:fig4} and Fig. \ref{fig:fig6}, We can see that the total utility has a significant reduced.

\subsection{The impact of weight parameters on the rate utility, reliability utility and network lifetime}
 In this section, we will investigate the impact of weight parameters $\gamma_s $  and $\varphi_s $  on the rate utility, reliability utility and network lifetime. The inherent tradeoff between rate utility and reliability utility can be observed from Fig. \ref{fig:fig9}, where $\gamma_s {\rm{ = }}1$ (i.e., network lifetime isn't taken into account), and  $\varphi_s $  ranges from 0 to 1. We can see that as $\varphi_s $  increases from 0 to 1, the rate utility increases, but the reliability utility decreases; and the larger the $\varphi_s $  , the larger drop in reliability utility.

Fig. \ref{fig:fig10} shows the inherent tradeoff between rate utility and network lifetime, where $\varphi_s {\rm{ = }}1$ (i.e., reliability isn't taken into account), and $\gamma_s $ ranges from 0 to 1. The network lifetime is shown in seconds. From Fig. \ref{fig:fig10}, we observe that as $\gamma_s $  increases from 0 to 1, the rate utility increases, while the network lifetime decreases and drops sharply when $\gamma_s $  is small. So, there is a evidently tradeoff between rate utility and network lifetime.

In the realistic applications, according to the actual requirements, we properly determine the value of weight parameters $\gamma_s $ and $\varphi_s $  to achieve a desired performance at the transmission rate, delivery reliability and network lifetime.

\section{Conclusions}
We have studied the rate-reliability-lifetime tradeoff in WSNs with link capacity constraint, reliability constraint and energy constraint. Our study is the first to jointly consider rate, reliability, and network lifetime in this rigorous tradeoff framework. Our optimization formulation for the rate-reliability-reliability tradeoff is neither separable nor convex. We convert the new formulation into a separable and convex optimization problem through a series of transformations.
A distributed SDD is developed for solving the rate-reliability-lifetime tradeoff problem. We also demonstrate that the convergence speed of SDD through numerical examples. Finally, We investigated the impact of different weight parameters $\gamma_s $  and $\varphi_s $  on the rate utility, reliability utility and network lifetime through numerical examples. We can select the appropriate value of weight parameters according to the actual requirements to achieve a desired performance of WSNs.

\ifCLASSOPTIONcaptionsoff
  \newpage
\fi

\begin{figure}[htbp]
\centering
\includegraphics[width=7.in]{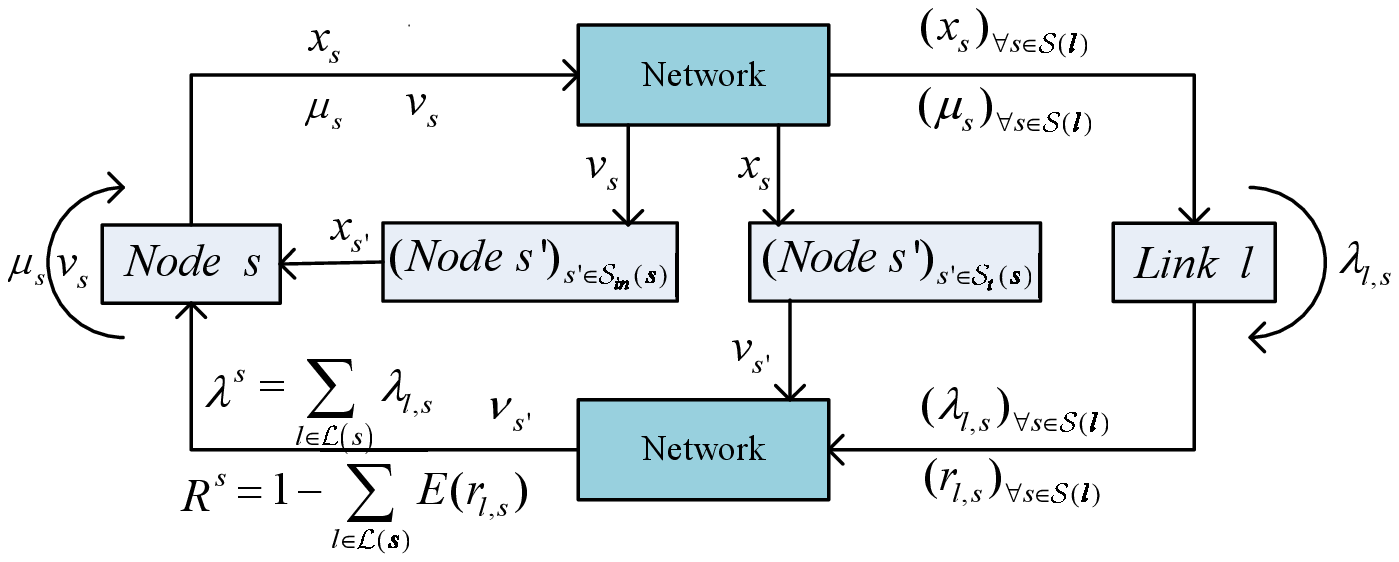}
\caption{Diagram for the distributed algorithm.}
\label{fig:fig1}
\end{figure}

\begin{figure}[htbp]
\centering
\includegraphics[scale=0.6,bb=88 366 495 765]{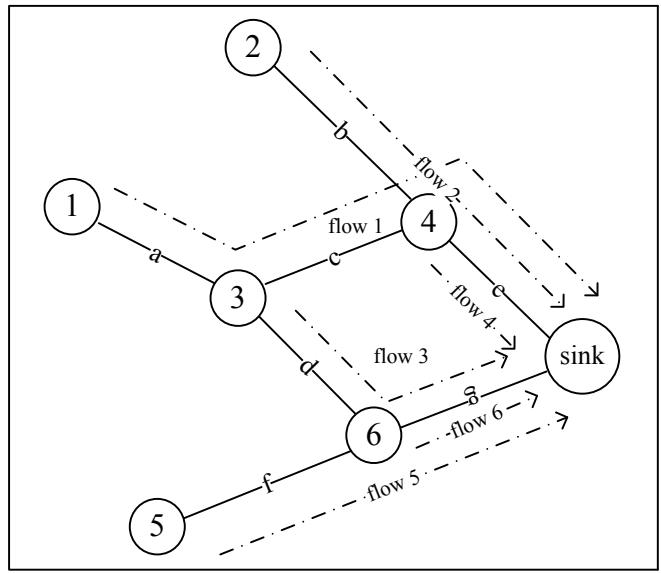}
\caption{Topology of wireless sensor network.}
\label{fig:fig2}
\end{figure}

\begin{figure}[t]
\centering
\includegraphics[scale=1,bb=105 465 473 755]{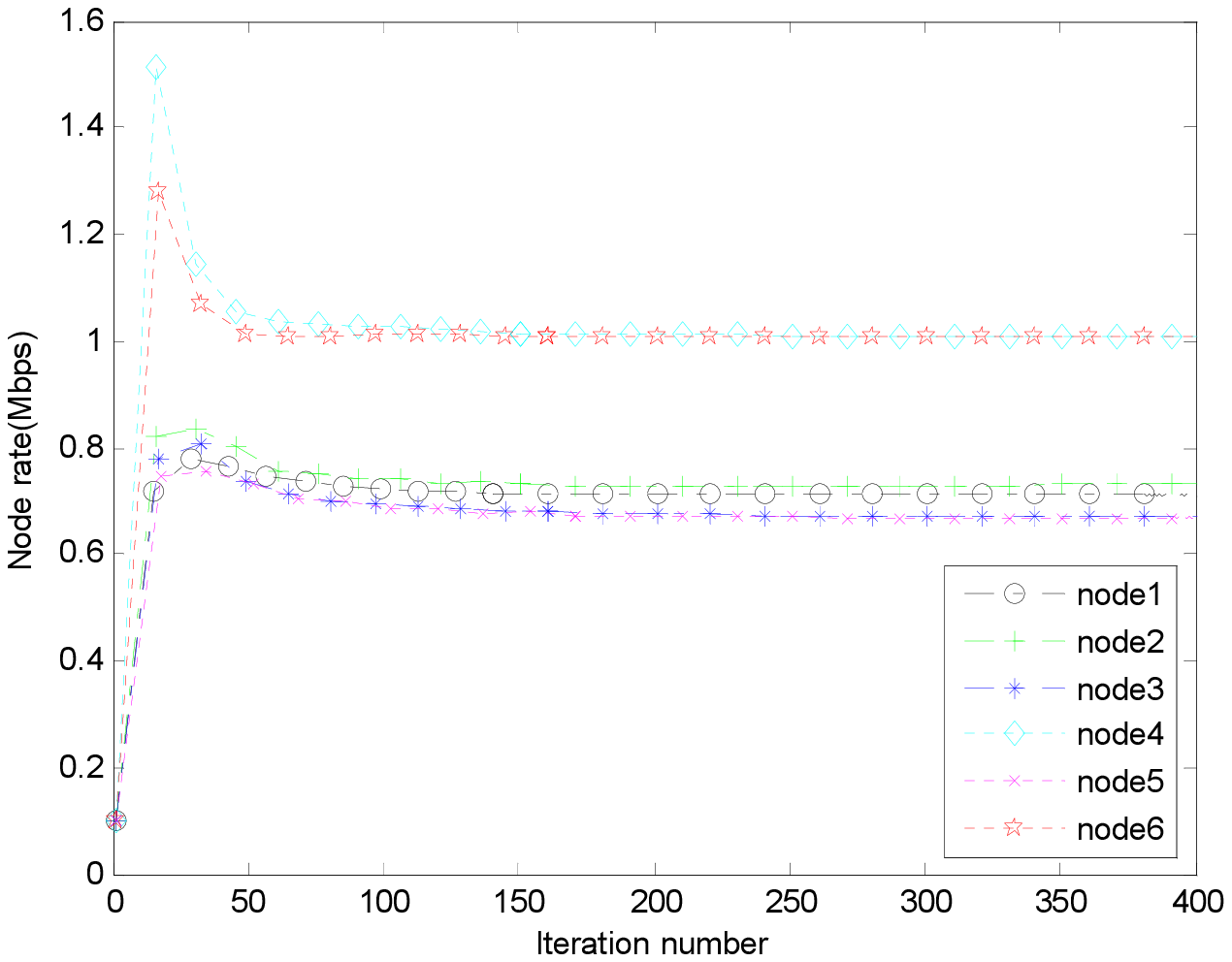}
\caption{The convergent performance of node rates in SDD with $\gamma_s {\rm{ = }}0.8$  and $\varphi_s {\rm{ = }}0.8$ .}
\label{fig:fig3}
\end{figure}

\begin{figure}[t]
\centering
\includegraphics[scale=1,bb=105 465 473 755]{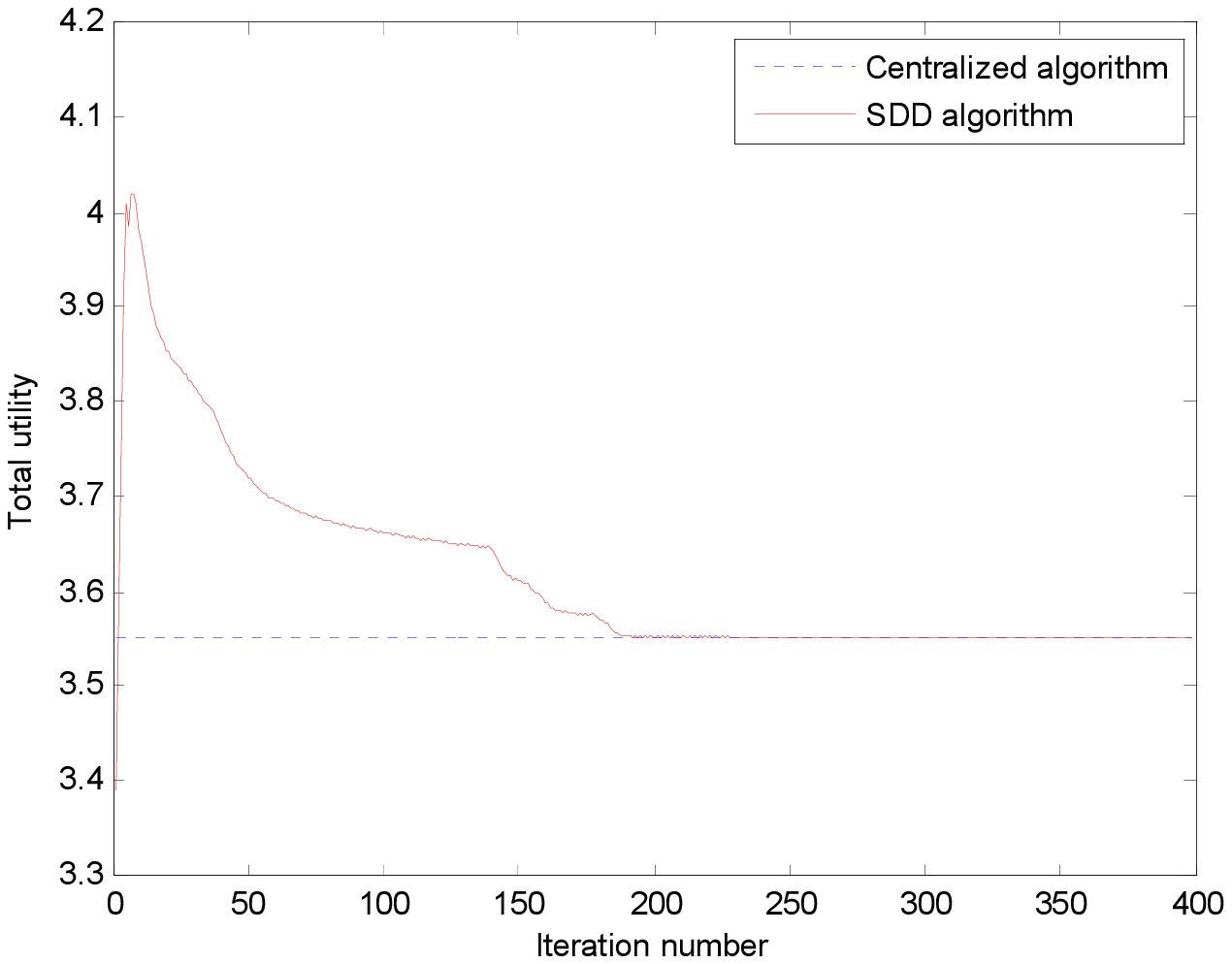}
\caption{The convergent performance of total utility in SDD with $\gamma_s {\rm{ = }}0.8$ , $\varphi_s {\rm{ = }}0.8$  .}
\label{fig:fig4}
\end{figure}

\begin{figure}[t]
\centering
\includegraphics[scale=1,bb=105 465 473 755]{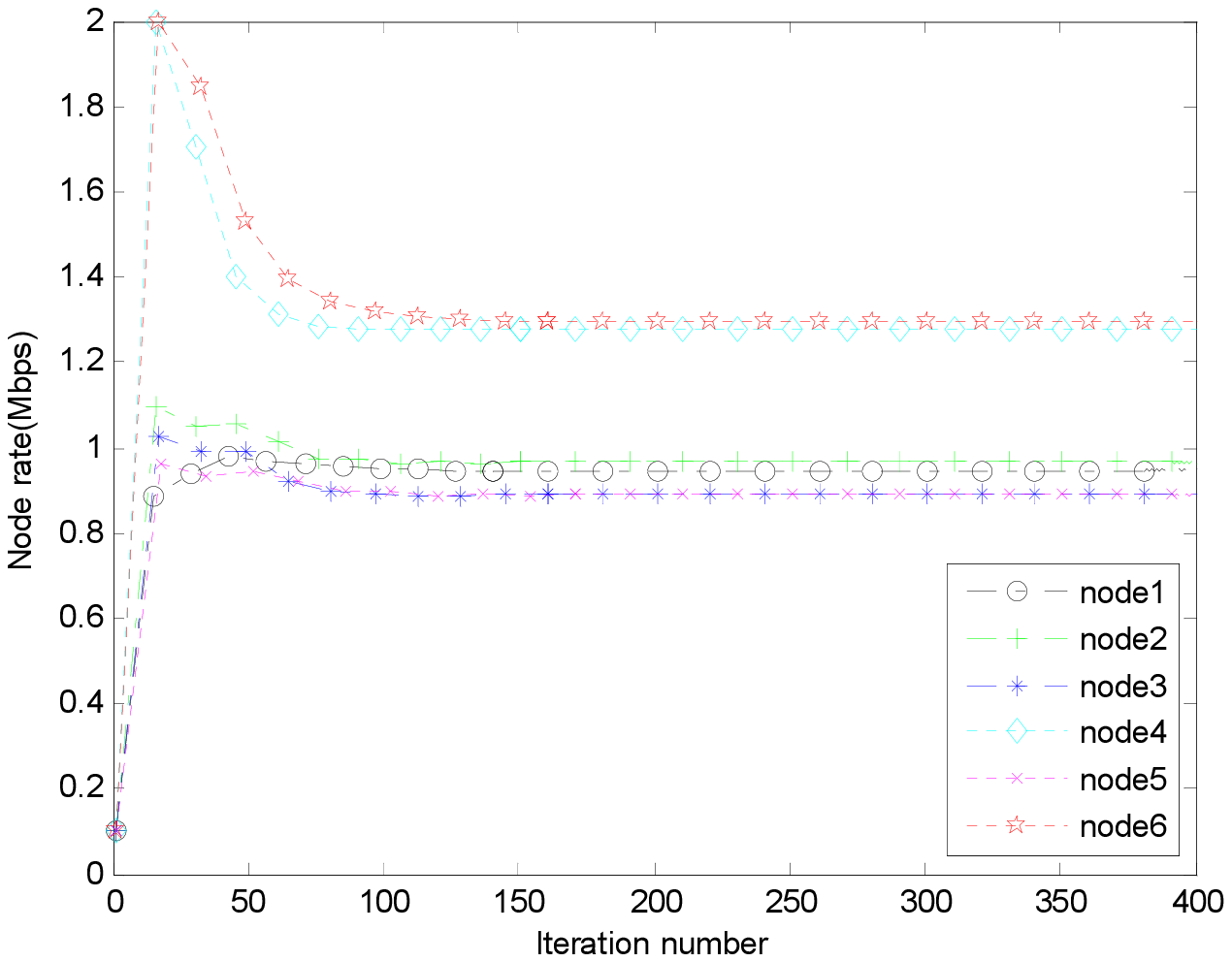}
\caption{The convergent performance of node rates in SDD with $\gamma_s {\rm{ = }}0.97$  and $\varphi_s {\rm{ = }}0.97$ .}
\label{fig:fig5}
\end{figure}

\begin{figure}[t]
\centering
\includegraphics[scale=1,bb=105 465 473 755]{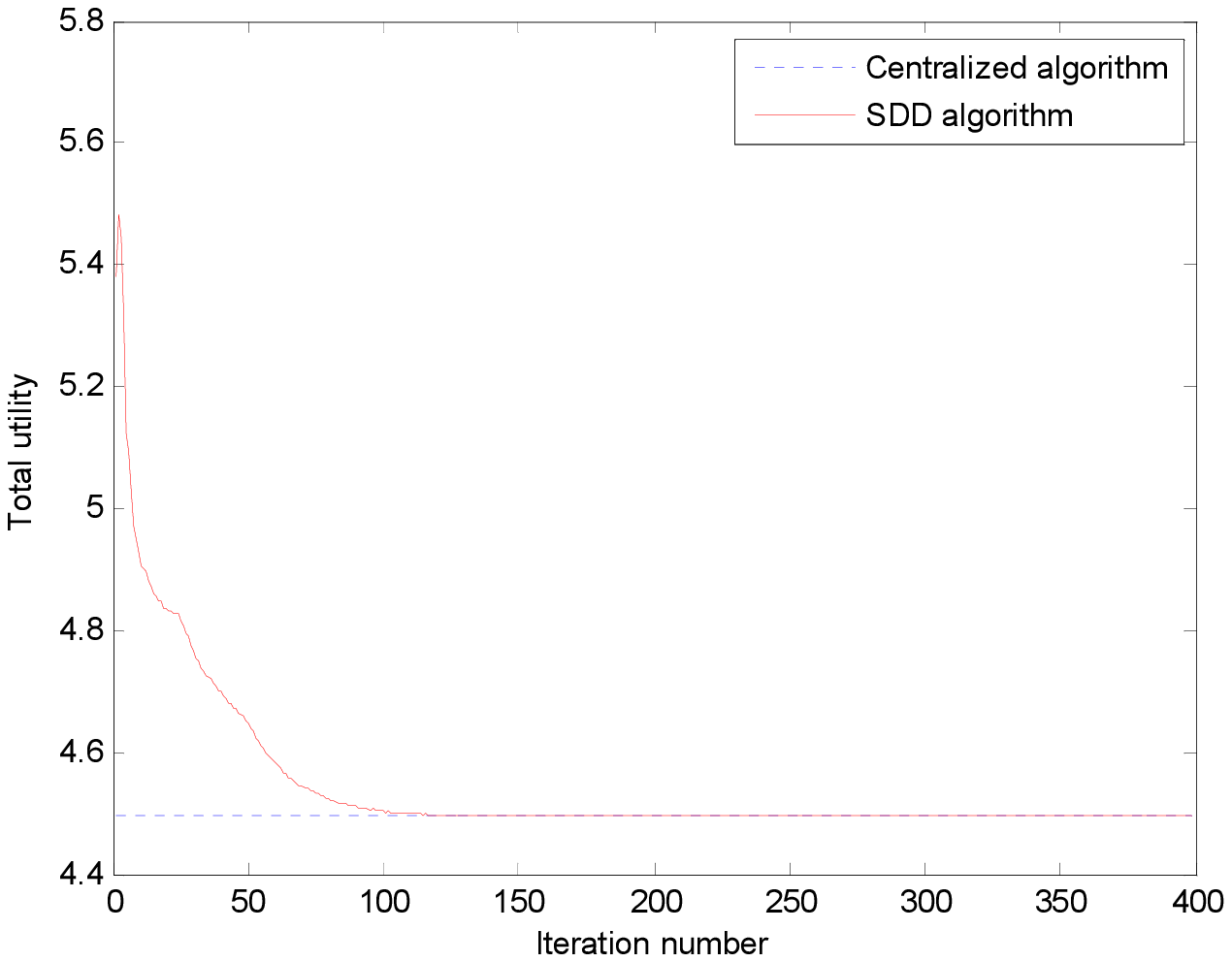}
\caption{The convergent performance of total utility in SDD with $\gamma_s {\rm{ = }}0.97$ , $\varphi_s {\rm{ = }}0.97$  .}
\label{fig:fig6}
\end{figure}

\begin{figure}[t]
\centering
\includegraphics[scale=1,bb=105 465 473 755]{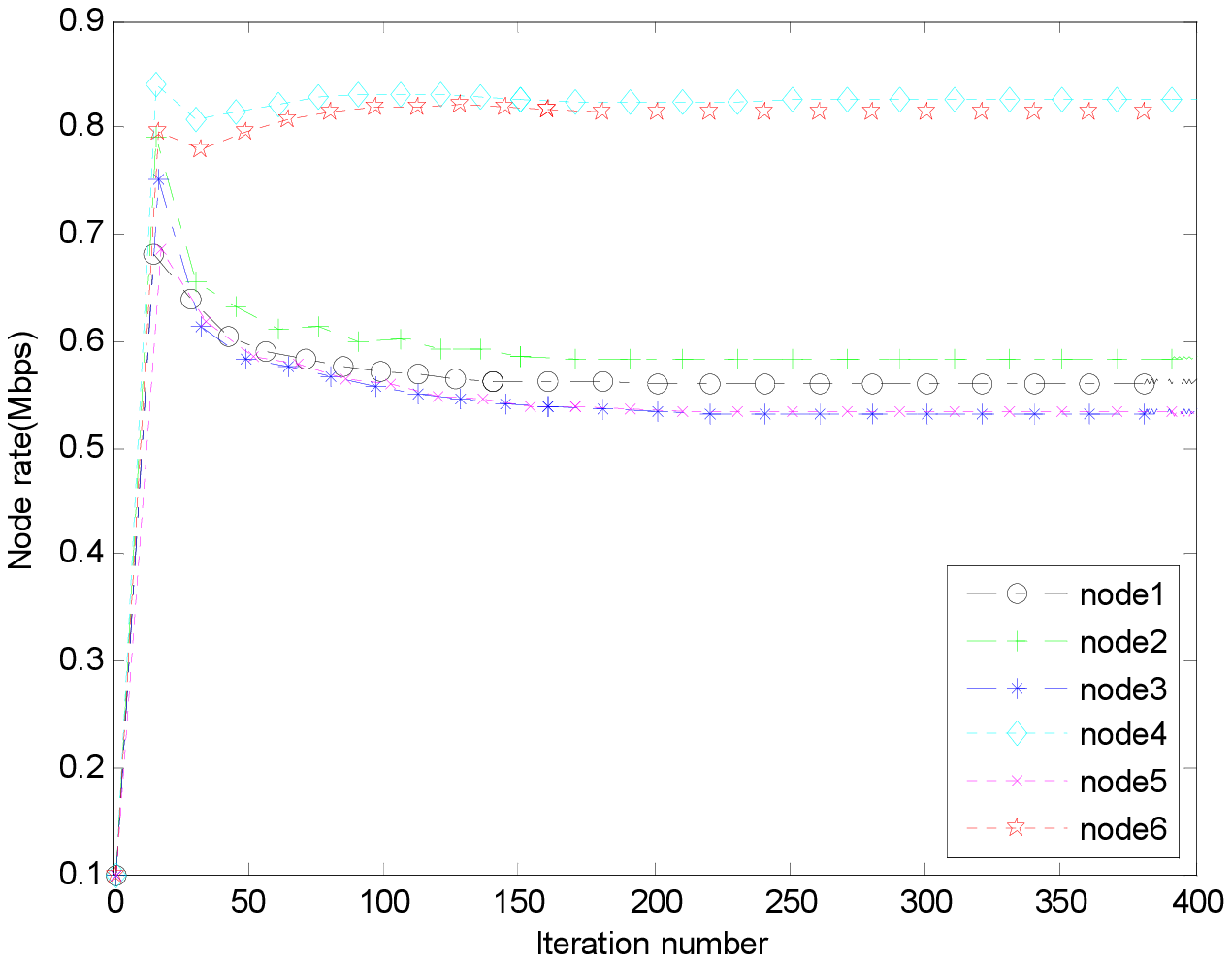}
\caption{The convergent performance of node rates in SDD with $\gamma_s {\rm{ = }}0.5$  and $\varphi_s {\rm{ = }}0.5$ .}
\label{fig:fig7}
\end{figure}

\begin{figure}[t]
\centering
\includegraphics[scale=1,bb=105 465 473 755]{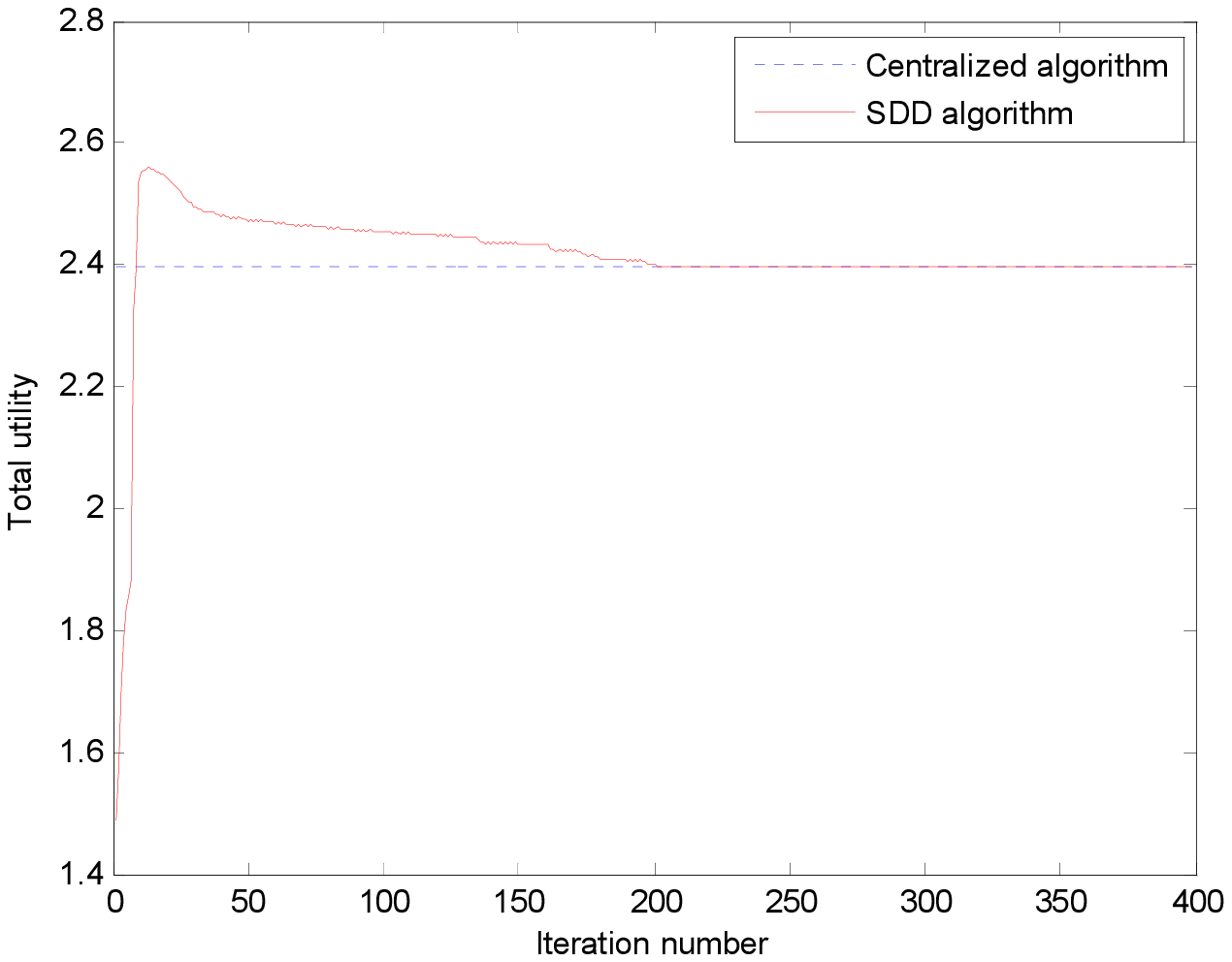}
\caption{The convergent performance of total utility in SDD with $\gamma_s {\rm{ = }}0.5$ , $\varphi_s {\rm{ = }}0.5$  .}
\label{fig:fig8}
\end{figure}

\begin{figure}[t]
\centering
\includegraphics[width=6in]{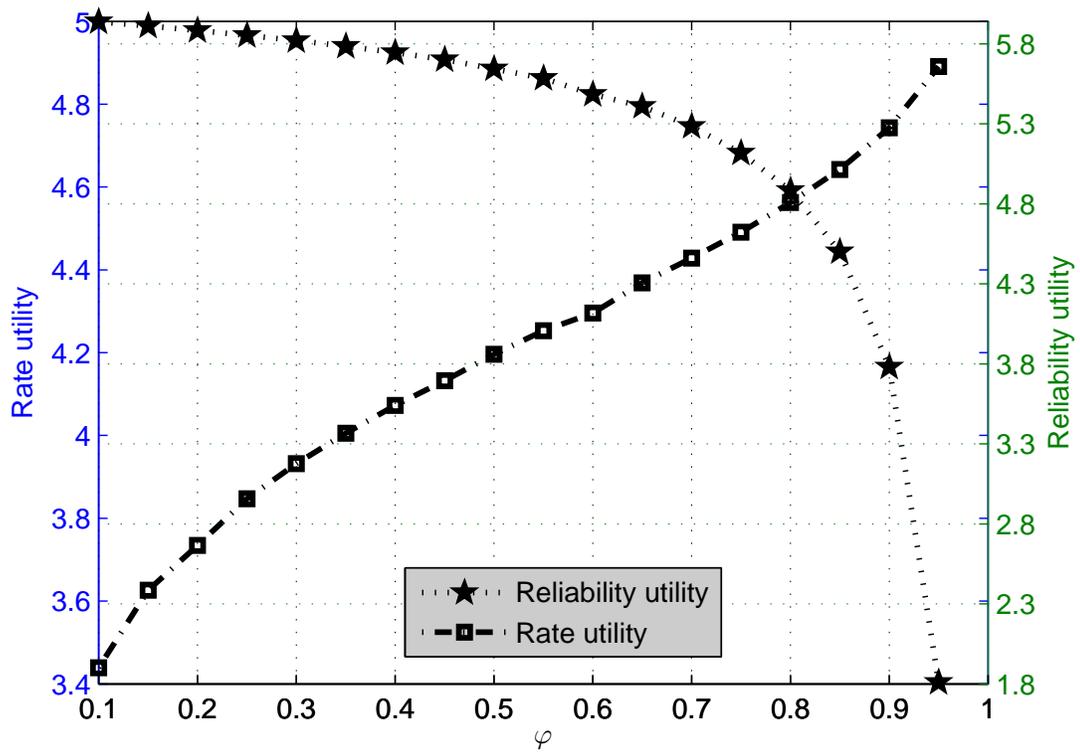}
\caption{The impact of weight parameter $\varphi_s $  on rate utility and reliability utility at $\gamma_s {\rm{ = }}1$ .}
\label{fig:fig9}
\end{figure}

\begin{figure}[t]
\centering
\includegraphics[width=6in]{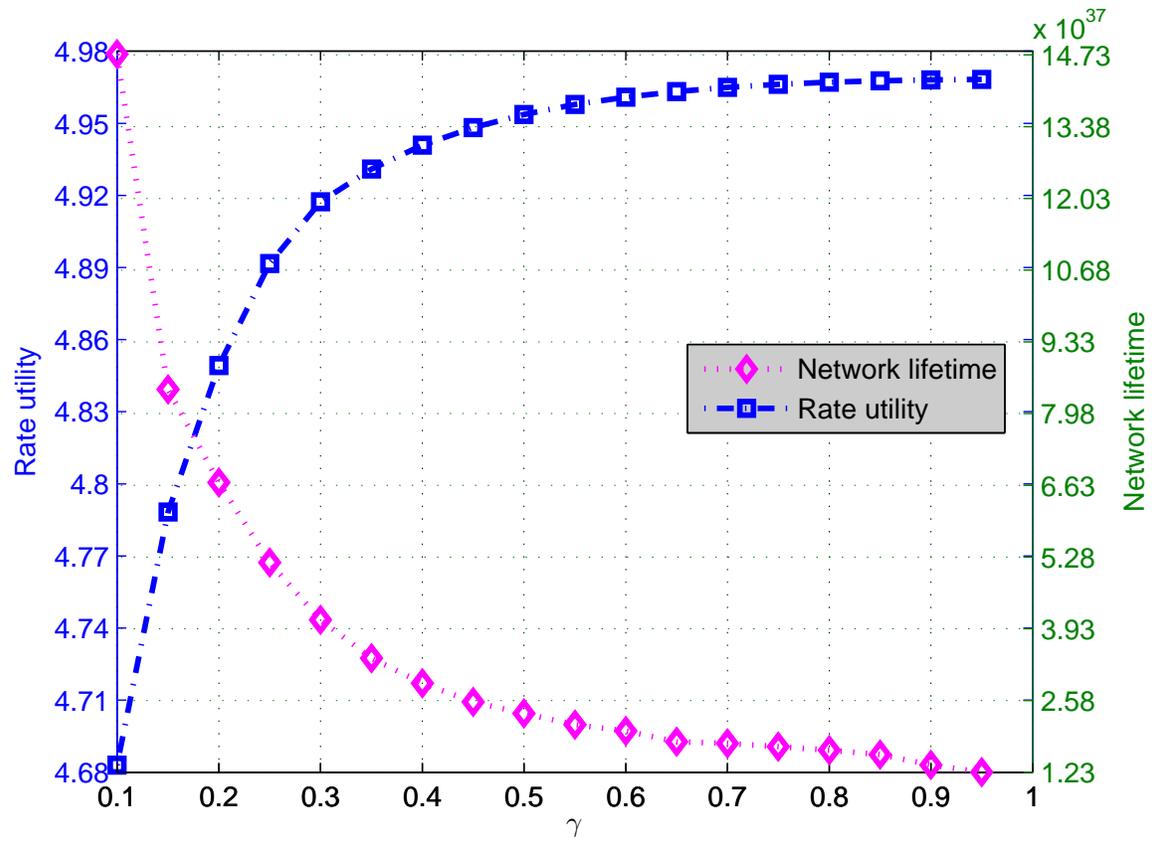}
\caption{The impact of weight parameter $\gamma_s $  on rate utility and network lifetime at $\varphi_s {\rm{ = }}1$ .}
\label{fig:fig10}
\end{figure}

\end{document}